\documentclass{article}
\usepackage{graphicx} 
\usepackage{caption}
\usepackage{amsmath}
\usepackage{amssymb}
\usepackage{amsthm}
\usepackage[table]{xcolor}
\usepackage{multirow}
\usepackage{makecell}
\usepackage{tikz}
\usepackage{svg}
\usetikzlibrary{positioning}
\usepackage{hyperref}
\usepackage{cleveref}
\usepackage{natbib}
\usepackage{authblk}
\usepackage[toc,page, title, titletoc]{appendix}

\crefname{appsec}{appendix}{appendices}
\Crefname{appsec}{Appendix}{Appendices}

\newcommand\independent{\protect\mathpalette{\protect\independenT}{\perp}}

\def\independenT#1#2{\mathrel{\rlap{$#1#2$}\mkern2mu{#1#2}}}
\usepackage[section]{placeins}

\newcommand{\cates}[0]{\hat{\tau}_1, \ldots, \hat{\tau}_K}

\usepackage{geometry}
\DeclareMathOperator{\E}{\mathbb{E}}

\title{Robust CATE Estimation Using Novel Ensemble Methods}

\author[1]{Oshri Machluf\thanks{oshri@phasevtrials.com}}
\author[1]{Tzviel Frostig}
\author[1]{Gal Shoham}
\author[1]{Tomer Milo}
\author[1]{Elad Berkman}
\author[1]{Raviv Pryluk}

\affil[1]{Research Department, PhaseV}

\begin{document}

\maketitle

\begin{abstract}
    The estimation of Conditional Average Treatment Effects (CATE) is crucial for understanding the heterogeneity of treatment effects in clinical trials. We evaluate the performance of common methods, including causal forests and various meta-learners, across a diverse set of scenarios, revealing that each of the methods struggles in one or more of the tested scenarios. Given the inherent uncertainty of the data-generating process in real-life scenarios, the robustness of a CATE estimator to various scenarios is critical for its reliability.
    To address this limitation of existing methods, we propose two new ensemble methods that integrate multiple estimators to enhance prediction stability and performance - \textit{Stacked X-Learner} which uses the X-Learner with model stacking for estimating the nuisance functions, and \textit{Consensus Based Averaging (CBA)}, which averages only the models with highest internal agreement. We show that these models achieve good performance across a wide range of scenarios varying in complexity, sample size and structure of the underlying-mechanism, including a biologically driven model for PD-L1 inhibition pathway for cancer treatment. Furthermore, we demonstrate improved performance by the Stacked X-Learner also when comparing to other ensemble methods, including R-Stacking, Causal-Stacking and others.
\end{abstract}

\section{Introduction}

Clinical trials aim to estimate the safety and efficacy of a treatment, typically in comparison to some control group. Efficacy is measured by one or more clinical outcomes, commonly referred to as the trial's \textit{Endpoints} (EP). The primary measure of interest in terms of efficacy is the \textit{Treatment Effect}, which is the expected difference in outcome between the treated patients and the control patients. 
Since only one potential outcome is observed for each patient, the treatment effect can be inferred by comparing the \textbf{average outcomes} of multiple patients—some given the treatment and some the control. If treatment assignment is independent of potential outcomes (i.e., the probability of treatment is unaffected by the expected treatment efficacy), the difference between the average outcomes of treated and control patients represents the Average Treatment Effect (ATE) \cite{rubin1974estimating}.
$$\E[Y^1 - Y^0] \stackrel{A \independent \{Y^0, Y^1\}}{=} \E[Y|A=1] - \E[Y|A=0],$$ where $A$ is the treatment assignment (0 for control, 1 for treatment) and $Y^a$ is the outcome under treatment assignment $a$. In clinical trials, the independence of treatment from potential outcomes is achieved by randomization.

There is often considerable variability in treatment effects among different patients \cite{mccormick2003geographic, kravitz2004evidence}, making ATE an insufficient indicator of treatment performance. In such cases, the \textit{Conditional Average Treatment Effect (CATE)} - the expected treatment effect, conditional on the patient's features, $X$ - is of great interest. Other parameters of the individual treatment effect distribution can be of interest as well, e.g., the proportion of treatment responders \citep{frostig2024causal}.

In recent years, CATE estimation has been extensively studied in various fields \citep{curth2021really, kunzel2019metalearners, wager2018estimation, jacob2021cate}, including the clinical field \citep{curth2024using}. Two main approaches that have gained popularity are Meta-Learners and Causal Forests. Meta-Learners decompose the problem of estimating the unobserved causal effect into sub-problems of predicting nuisance functions: the different potential outcomes and the propensity. Meta-Learners can leverage any Machine-Learning (ML) model (called base-learners) to estimate the nuisance functions. Several different frameworks were suggested on how to combine these different estimates to obtain a CATE estimator (including S-Learner, T-Learner, X-Learner and R-Learner which are all described in \cref{sect:related_work}, among others). Causal Forests generalize Random-Forests for the task of CATE estimation by adapting the evaluation and splitting criteria to consider the expected difference in outcomes between treatments instead of the expected outcome. 

In real-life applications, the underlying Data Generation Process (DGP) is unknown. Additionally, because CATE cannot be directly observed, standard cross-validation methods used in supervised learning settings are inapplicable. This challenge has led to numerous studies on how to select the best CATE estimator by examining various loss functions. 

\cite{schuler2018comparison} found that a loss function based on the R-Learner performs best for CATE selection. \cite{mahajan2022empirical} extended this work, finding that with an Auto-ML approach to the base-learners, the T-Learner and X-Learner based losses outperform the R-Learner based loss. \cite{doutreligne2023select} recommended using the R-Loss for model selection, stacking for nuisance function estimation, and a 90/10\% split for model estimation and selection. \cite{curth2024using} compared selection metrics based on plug-in estimates, factual estimation, and pseudo-outcomes, arguing that no single metric is universally recommended as it depends on the dataset. However, they recommend the pseudo-outcome as the default approach.

An alternative approach to selecting the best CATE estimator involves averaging multiple CATE estimators rather than choosing a single one. \cite{nie2021quasi} proposed R-Stacking, which combines several CATE estimators by finding the optimal linear combination that minimizes the R-Loss, leveraging the strengths of each individual estimator. \cite{han2022ensemble} introduced Causal-Stacking for randomized experiments, obtaining a linear combination of CATE estimators that minimizes the mean squared error of a pseudo outcome based on the , with non-negative weights that sum to one. It was shown to outperform R-Stacking in balanced treatment allocation settings.  

In this work our goal is to study CATE estimation using ensembles in clinical trial data. We designed our evaluation process to emphasise the unique challenges and characteristics of such datasets. In addition to using various additive models based on combinations of linear and nonlinear transformations of features, we also used a biologically driven DGP, which utilises a graphical functional model built to describe the Programmed-Death-Ligand 1 (PD-L1) inhibitor in cancer treatment. A distinct characteristic of clinical trials, especially in Phase II exploratory trials, is their relatively small sample sizes, which makes it challenging to learn complex relations from data. Furthermore, it makes it difficult to utilize techniques involving sample splitting. Therefore, we focus on small datasets, ranging from 100 to 750 patients, representing the setting of interest. Current benchmarks of CATE estimation usually consider a higher number of observations \citep{schuler2018comparison, jacob2021cate, mahajan2022empirical, acharki2023comparison, doutreligne2023select}. Only \cite{nie2021quasi, curth2024using} considered data-sets with 500 observations. 

Furthermore, we propose two new ensemble methods for CATE estimation - \textbf{Stacked X-Learner} which is an application of the X-Learner Meta-Learner using stacked ML predictors for the nuisance functions, and the \textbf{Consensus Based Averaging (CBA)} which compares several CATE estimates, selects a subset of models that yield the estimates with highest internal agreement and averages them. Both methods are described in \cref{sect:suggested_ensembles}. 

One of the main tasks in the analysis of clinical data is \textit{Subgroup identification} \cite{lipkovich2017tutorial} - the process of identifying a suitable target population for a treatment.
We hypothesize that good CATE estimation can be beneficial for guiding the subgroup identification process. In order to keep the focus of this paper on CATE estimation, we leave an in-depth analysis of model-based subgroup search to future work, and only propose a designated evaluation method for CATE estimators, which highlights their accuracy on potential subgroups.

In the next section we introduce the relevant notation and terminology. \Cref{sect:related_work} overviews different CATE estimators and ensembles of CATE estimators.  In \cref{sect:suggested_ensembles} we propose our two ensemble methods. In \cref{sect:dgp} we describe the data-generation-process (DGP) used in the simulation, \cref{sect:metrics} is devoted to the evaluation metrics. In \cref{sect:simulation} we present our simulation study and results. Future work will be dedicated to other common endpoint types.

\section{Notation and Terminology}\label{sect:notation}

We Assume each patient $i$ can be described using:
\begin{itemize}
    \item $x_i$ - a vector of patient features/covariates. We assume the dimension of the feature space is $P$ and denote the $p$-th feature of the $i$-th patient as $x^p_i$.
    \item $A_i, a_i$ - treatment assignment/allocation. We limit the discussion in this paper to binary treatments, $a_i=0$ indicates assignment of patient $i$ to the control and $a_i=1$ indicating assignment to treatment.
    \item $\pi(x_i)$ - the propensity score of each patient assuming strong ignorability, then $\pi(x_i) = \mathbb{P}(a_i = 1)$.
    \item $y_i$ - the patient's measured EP value. We refer to this also as outcome/response interchangeably. $y_i$ is a random function of $x_i$ and $a_i$.
\end{itemize}
Adopting the potential outcomes framework, we also assume each patient has:
\begin{itemize}
    \item $Y^0_i$, $Y^1_i$ - the patient's hypothetical response under control or treatment, respectively. For control patients $y_i = y^0_i$ and for treated patients $y_i = y^1_i$.
    \item $\tau_i = y^1_i - y^0_i$ - the patient's \textit{Individual Treatment Effect (ITE)}. This is an unobserved quantity, since we can only observe one potential outcome for each patient.
    \item $\tau(x_i) = \E[Y^1 - Y^0 | X=x_i]$ - the \textit{Conditional Average Treatment Effect} based on the patients features. By definition $\tau(x) = \E[\tau|X=x]$.
\end{itemize}

We denote estimators using $\hat{\quad}$, e.g., a CATE estimator is denoted by $\hat{\tau}$.  We follow the framework used by \cite{lipkovich2017tutorial} for decomposing $Y$ into \textit{prognostic} and \textit{predictive} components (with some changes in notation for internal consistency):
$$\E[Y|X, A] = \gamma(X) + \tau(X)\cdot A,$$
with $\gamma(X)$ being the prognostic function, which describes the expected outcome under control, and $\tau(X)$ (CATE) being the predictive factor - the expected difference between outcomes under treatment vs. under control.

\section{Related Work} \label{sect:related_work}

This analysis uses two general types of estimators - Causal Forests \citep{athey2016recursive, wager2018estimation} and Meta-Learners \citep{kunzel2019metalearners}. For each, we have explored a wide range of possible configurations with the aim of reaching robust performance across scenarios. 

\textbf{\textit{Causal Forest (CF)}} is an adaptation of random forests - an average of regression trees. In causal forest, the base trees composing the forest are \textit{causal trees}, aimed at estimating local differences between average potential outcomes \cite{athey2016recursive}. To handle issues of selective inference they introduced the notion of honesty, where the sample is split into a training and estimation set. The training set is used to identify the partitions of the tree and the estimation set to estimate the CATE within the partition. Furthermore, adapting ideas of doubly robust estimation for CATE \cite{chernozhukov2017double}, CF also has debiased implementations for mitigating potential bias caused by non-random treatment allocation \cite{wager2018estimation}. 

\textbf{\textit{Meta-Learners}} are an estimation framework for treatment effect, that enables using any ML model as a “base learner” for learning various nuisance functions and composing an estimator for CATE using a transformation of the learned functions \cite{kunzel2019metalearners, kennedy2023towards}. There are several common meta-learner structures, out of which we compared the following:
\begin{itemize}
    \item \textbf{S learner}: A single model is trained to regress the outcomes on the features and the treatment assignment (the treatment is modeled as an additional binary variable attached to $X$).
    $$\hat y(x, a)=\hat\E [Y(X,A)|X=x, A=a]$$
    Then, ATE is estimated by contrasting this model's predictions for both potential outcomes:
    $$\hat \tau(x)=\hat y(x,1)-\hat y(x, 0)$$
    \item \textbf{T learner}: This approach uses base-learners to estimate the conditional expectations of the two potential outcomes separately - $\{(X, Y)\ ;\  A=0\}$ are used to train $\hat\mu_0(X)$, an estimator for $\E[Y|A=0, X]$ and $\{(X, Y)\ ;\ A=1\}$ to train $\hat\mu_1(X)$ - an estimator for $\E[Y|A=1, X]$. Finally, an estimator for CATE is obtained by subtracting them: 
    $$\hat\tau(X)=\hat\mu_1(X) - \hat\mu_0(X)$$
    \item \textbf{X learner}: This approach builds on the foundations of the T Learner, and starts similarly by estimating $\hat{\mu_0}(X)$ and $\hat{\mu_1}(X)$.
    it then uses these estimators to impute the missing potential outcomes and generate "pseudo individual effects":
    $$if\ A=0\ :\ \ D^0 := \hat\mu_1(X) - Y,$$
    $$if\ A=1\ :\ \ D^1 := Y - \hat\mu_0(X)$$
    Next, $D^0$ and $D^1$ are used to train two separate estimators for CATE - $\hat\tau_1(X) = \hat\E[D^1|X]$ and $\hat\tau_0(X) = \hat\E[D^0|X]$.
    Finally, a weighted average of the two estimates is used to estimate CATE. a common weighting choice is the propensity, which is also estimated from the data:
    $$\hat\tau(X) = \hat\pi(X)\cdot\hat\tau_0(X) + (1-\hat\pi(X))\cdot\hat\tau_1(X).$$

    \item \textbf{R learner}: This approach suggests to minimize the following loss function, \begin{equation*}
    \tau(\cdot) = \arg\min_{\tau} \left\{ E \left( \left[ \{ Y_i - m(X) \} - \{ A_i - \pi(X) \} \tau(X) \right]^2 \right) \right\},
    \end{equation*} where $m(X_i) = \mathbb{E}(Y_i | X_i) = \mu_0(X_i) + \pi(X)\tau(X)$. In practice, the nuisance functions $m(X), \pi(X)$ are estimated in a cross-fitting manner.
    
    \item \textbf{DR learner}: This approach constructs a doubly-robust pseudo-outcome for CATE using a sub-sample of the training data, and uses the rest of the train-set to regress this pseudo outcome on X. First, using the first subset $S_1$ to train $\hat\pi(X),\ \hat\mu_0(X),\ \hat\mu_1(X)$ - estimators for $\pi(X),\ \E[Y|A=0, X],\ \E[Y|A=1,X]$, respectively.
    Then, construct the following pseudo-outcome:
    $$\hat\varphi(X, Y, A)= \frac{A - \hat\pi(X)} {\hat\pi(X)\cdot(1 - \hat\pi(X))} (Y - \hat\mu_A(X)) + \hat\mu_1(X) - \hat\mu_0(X).$$
    Finally, use the rest of the train-sample, $S_2=S \setminus S_1$ to regress:
    $$\hat\tau_{dr}(x)=\hat\E[\hat\varphi(X, Y, A)|X=x].$$
\end{itemize}
Each of the aforementioned methods can use any ML model as a "base learner" for any of the nuisance functions it estimates. The choice of base estimators should account for the assumed complexity of the underlying data-generating process, the size of available training data etc. Clinical trials often test complex mechanisms, which favor more flexible ML models. However, the typical sample sizes in this setting are usually quite limited, favoring simpler more regularized models. 

\subsection{Ensemble Methods}

We will assume $K$ CATE estimation models denoted by $\cates$. Our goal is to find a function of the estimators which results in a single ensemble estimator, denoted by $\hat{\tau}^*$. 
\begin{itemize}
    \item \textbf{R-Stacking}: R-Stacking is based on finding the best linear combination of $\cates$ which minimizes the R-loss \citep{nie2021quasi},
    \begin{equation*}
    (\hat{b}, \hat{c}, \hat{\alpha}) = \arg\min_{b, c, \alpha} \left\{ \sum_{i=1}^{n} \left[ \left\{ Y_i - \hat{m}(X_i) \right\} - b - \left\{ c + \sum_{k=1}^{K} \alpha_k \hat{\tau}_k(X_i) \right\} \left\{ A_i - \hat{\pi}(X_i) \right\} \right]^2 \right\},
    \end{equation*} subject to $\alpha_k > 0, \forall k$. $n$ is the number of observation in the validation set. The CATE estimation according to the ensemble is $$\hat{\tau}^*(x) = \hat{c} + \sum_{k=1}^{K} \alpha_k \hat{\tau}_k(x). $$

    \item \textbf{Causal Stacking}: \cite{han2022ensemble}, focused on the setting of a randomized experiment, i.e., the true propensity, $\pi(x)$ is known.  Their suggestion is to obtain a linear combination of estimators which minimizes the mean square error of a pseudo outcome, \begin{equation*} \hat{\alpha} = \arg\min_{\alpha} \left\{ \sum_{i=1}^{n} \left[  \hat{\tau}_i^{\mathrm{PO}} - \sum_{k=1}^{K} \alpha_k \hat{\tau}_k(X_i) \right]^2 \right\}, \end{equation*} subject to $\sum_{k=1}^K \alpha_k = 1$ and $\alpha_k > 0, \forall k$. The psuedo outcome is defined by, $$\hat{\tau}_i^{\mathrm{PO}} = \left( \hat{\mu}_1(X_i) - \hat{\mu}_0(X_i) \right)  + \frac{ \left( Y_i - \hat{\mu}_1(X_i) \right) A_i}{{\pi}(X_i)}  - \frac{ \left( Y_i - \hat{\mu}_0(X_i) \right) \left( 1 - A_i \right)}{1 - {\pi}(X_i)}.$$ Causal-Stacking was shown to perform better than the R-Stacking when the treatment allocation is balanced. Both methods require that the estimated nuisance functions, $\hat{m}, \hat{\pi}, \hat{\tau_{k}}$ or $\hat{\mu}_1, \hat{\mu}_0$ are estimated on a different set then the one used for obtaining $\hat{\alpha}_k$
    \item \textbf{T-Stacking}: The ideal method to train an ensemble, would be to regress the resulting CATE models estimates against the true CATE of a validation set. Unfortunately, we can only estimate CATE and not directly observe it. To circumvent the issue, it was suggested to replace the true CATE with some CATE estimator trained on the validation set.  The T-score, is one such loss, where the CATE is estimated using a T-Learner \citep{alaa2019validating, mahajan2022empirical}.  T-Stacking is an adaptation of the T-score for the purposes of ensembeling,  $$\hat{\alpha} = \arg \min_{c, \alpha} \left\{ \sum_{i=1}^n \left[ \hat{\tau}_T(X_i) - \sum_{k=1}^K\alpha_k \hat{\tau}_k \right]^2 \right\}$$ subject to $\alpha_k > 0$. Here $\hat{\tau}_T$ is a T-Learner trained on the validation-set. 
  \end{itemize}

\section{Suggested Ensembles} \label{sect:suggested_ensembles}

We suggest two novel methods of ensembling. One, motivated by the work of \cite{doutreligne2023select}, 
is based on the X-Learner and stacking for the estimation of the nuisance functions. The second method is based on finding consensus among models and does not require sample splitting to obtain the ensemble estimator. This method is motivated by the need to efficiently use the limited sample size available in clinical trials.

\begin{itemize}
    \item \textbf{Stacked X-Learner}: Motivated by the results of \cite{doutreligne2023select}, we suggest an X-Learner based on stacking. We use stacking for the first level ($\mu_0$, $\mu_1$) and second level ($\tau_0$, $\tau_1$) of nuisance functions. Each function is estimated as a standard supervised learning task, making it feasible to use the typical stacking methods.
    \item \textbf{Consensus Based Averaging (CBA)}: Existing ensemble methods are not sample efficient; they require splitting the data into a training set, where the nuisance functions and the CATE are estimated, and a validation (or averaging) set, which is used to train the ensemble estimator. This method prioritizes models with higher agreement. Initially, predictions from each model are collected. Kendall's Tau correlation coefficients are calculated between each pair of models, and the mean correlation coefficient for each model is determined. Models are then sorted based on their mean correlation values. The method identifies a subset of models that are in higher agreement by detecting the 'knee' in the sorted mean correlation curve. Models up to this knee point are selected, and their predictions are averaged to produce the final ensemble prediction, thereby enhancing robustness and accuracy. For further details on the method, see \cref{appendix:cba}.    
\end{itemize}

\section{Data Generation Process} \label{sect:dgp}

We considered three groups of DGP. The first uses a functional graphical model \citep{scholz2013explanation} to concisely model the PD-L1 pathway in urothelial cancer. The second is a flexible additive linear model with nonlinear transformations and second order interactions, which allows to easily create diverse and tunable settings.
The third DGP is based on the ACIC2016 datasets \cite{dorie2019automated}, which use a combination of real-world variable data and synthetic specified treatment effects.

\subsection{PD-L1-biological data generation model}
\label{appendix:pdl1}
This DGP aims to mimic a biological process related to the progression of tumors, a setting often encountered in clinical trials. The DGP employs a mechanistic disease model that forms a concise description of a main pathway or mechanism of action for the progression of a selected disease. Such models require in-depth biological knowledge and are costly to construct, but they can generate more realistically distributed datasets with complex interactions formed by the mechanistic relations defined in the model.
We utilized a simplified mechanistic model for the \textbf{PD-L1 pathway in urothelial cancer} \citep{mariathasan2018tgfbeta}. This model describes the  connections between a few key factors related to the progression of urothelial tumors and specifically to their treatment using PD-L1 inhibitors.

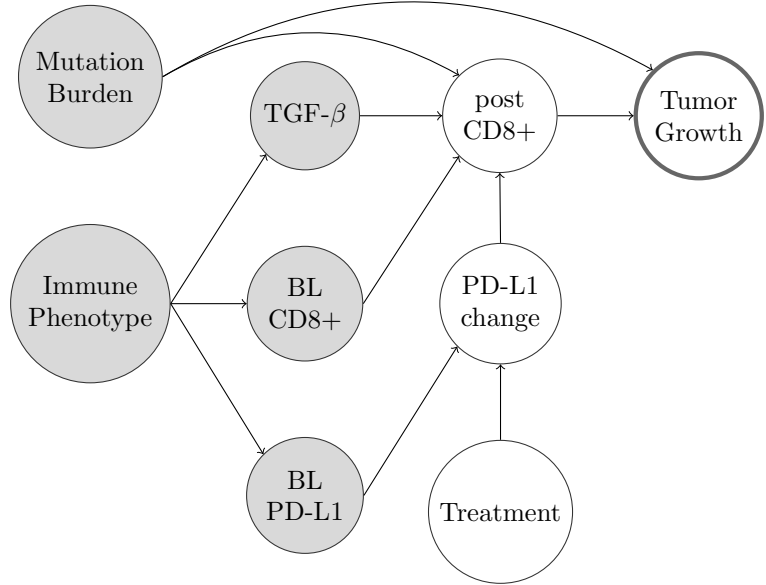
\begin{figure}[!htb]
\begin{center}
\begin{tikzpicture}[
baseline_node/.style={circle, draw=black!80, fill=gray!30, thin, minimum size=7mm, align=center},
interim_node/.style={circle, draw=black!80, thin, minimum size=7mm, align=center},
target_node/.style={circle, draw=black!60, ultra thick, minimum size=5mm, align=center},
]
\node[baseline_node] (I)                        {Immune\\Phenotype};
\node[baseline_node] (Epre)  [right=of I]       {BL\\CD8+};
\node[baseline_node] (Lpre)  [below=of Epre]    {BL\\PD-L1};
\node[baseline_node] (Bpre)  [above=of Epre]    {TGF-$\beta$};
\node[baseline_node] (M)     [above=of I]       {Mutation\\Burden};
\node[interim_node]  (dL)    [right=of Epre]    {PD-L1\\change};
\node[interim_node]  (Epost) [right=1.08 of Bpre]    {post\\CD8+};
\node[interim_node]  (T)     [below=of dL]      {Treatment};
\node[target_node]   (G)     [right=of Epost]   {Tumor\\Growth};

\draw[->] (I.east) -- (Bpre.south west);
\draw[->] (I.east) -- (Lpre.north west);
\draw[->] (I.east) -- (Epre.west);
\draw[->] (Bpre.east) -- (Epost.west);
\draw[->] (Epre.east) -- (Epost.south west);
\draw[->] (Lpre.east) -- (dL.south west);
\draw[->] (T.north) -- (dL.south);
\draw[->] (dL.north) -- (Epost.south);
\draw[->] (M.east) to [bend left] (Epost.north west);
\draw[->] (M.east) to [bend left] (G.north west);
\draw[->] (Epost.east) -- (G.west);

\end{tikzpicture}
\captionsetup{justification=centering}
\caption{\textbf{PD-L1 Model}, relating some key factors for tumor progression\\Shaded nodes compose the features $X$. Tumor Growth (thick circle on the right) is the endpoint $y$.}
\label{fig:pdl1}
\end{center}
\end{figure}

Figure \ref{fig:pdl1} illustrates the PD-L1 (Programmed Death Ligand 1) model, which delineates prominent factors influencing tumor progression. This model integrates various biological and immunological elements to simulate some key aspects of the dynamics of tumor growth.
\begin{itemize}
    \item \textbf{Mutation Burden}: One of two "root" elements in our model, this factor impacts both the post treatment CD8+ levels, as well as potentially impacting Tumor Growth in additional pathways not modeled here, thus represented as a direct impact on Tumor Growth.
    \item \textbf{Immune Phenotype}: The other "root" element, this ordinal variable classifies the presence of immune cells in the tumor area into 3 classes - "inflamed", "excluded" or "desert" - indicating decreasing levels of immune activity. Higher levels of activity act to increase  TGF-$\beta$ (Transforming Growth Factor-beta), basleine CD8+ levels and PD-L1 levels.
    \item \textbf{TGF-$\mathbf{\beta}$}: Causes the CD8+ levels to increase, regardless of the treatment.
    \item \textbf{CD8+ Effector}: We model the CD8+ levels post-treatment as the baseline levels, with the addition of terms relating to TGF-$\beta$ and to an interaction between mutation burden, and the change in PD-L1 caused by the treatment.
    \item \textbf{PD-L1}: We model the ligand levels as an ordinal factor with 3 levels. The model assumes a 1 level decrease with treatment.
    \item \textbf{Tumor Growth}: Tumor growth is driven by the mutation burden (increasing the growth), while it is inhibited by the post-treatment levels of CD8+ effector cells.
        
\end{itemize}

\subsection{Linear Models}
The second type of DGP is based on a linear model with the addition of nonlinear transformations and second order interactions between features.
\begin{equation}\label{eq:dgp}
\begin{split}
Y = \sum\limits_{j}{\beta^jf^j(X^j)}\ + \sum\limits_{j, k}{\beta^{jk}f^j(X^j)f^k(X^k)}\ + \\
\left({\sum\limits_{j}{\delta^jg^j(X^j)}\ + \sum\limits_{j, k}{\delta^{jk}g^j(X^j)g^k(X^k)}}\right)\cdot A +\ \epsilon,
\end{split}
\end{equation}
where $\beta^j$, $\beta^{jk}$, $\delta^j$ and $\delta^{jk}$ are constants, $f^j$ and $g^j$ nonlinear transformations of feature $X^j$, and $\epsilon$ a Gaussian random variable.\\
This approach allows simple control over the level of complexity of the signal and signal to noise ratios. Additionally, due to its linear form, its interpretation is relatively intuitive. We present the results for 6 representative scenarios based on this approach, consisting of three “types”:
\begin{itemize}
\item \textbf{Linear}: Scenarios that include linear prognostic and predictive components. i.e. a DGP of the form: $$Y = \sum\limits_{p \in \{1, 2.. P\}}{\beta^pX^p}\ + \sum\limits_{p \in \{1, 2.. P\}}{\delta^pX^p}\cdot A+\ \epsilon.$$
\item \textbf{Slightly nonlinear}: These scenarios contained nonlinear transformations for the predictive component, with no second-order interactions, resulting in the following form: $$Y = \sum\limits_{p \in \{1, 2.. P\}}{\beta^pX^p}\ + \sum\limits_{p \in \{1, 2.. P\}}{\delta^pg^p(X^p)}\cdot A+\ \epsilon.$$
\item \textbf{Highly nonlinear}: Scenarios employing the full complexity of the model presented in \ref{eq:dgp}, allowing for non-linearities in both prognostic and predictive features, including second order interactions.
\end{itemize}

\subsection{ACIC 2016 Datasets}

The ACIC 2016 datasets consist of real-world covariates, with simulated treatment and response. Using a mixture of real-world and simulated data provides an appealing compromise between maintaining the realistic complex structure of clinical data, and still having full knowledge of potential outcomes and their expected values. 

The covariates are taken from a longitudinal study aimed at identifying the causal covariates responsible for developmental issues \cite{dorie2019automated}. From the large dataset, 4802 observations and 58 covariates were selected.  Of these covariates, 3 are
categorical, 5 are binary, 27 are count data, and the remaining 23 are continuous.

Several datasets are created using the covariates, representing a large variety of treatment allocation mechanisms, prognostic and predictive factors \cite{dorie2019automated}. These datasets are  frequently referenced and used in other comparative analyses, e.g. \cite{mahajan2022empirical, han2022ensemble, curth2023search}.
Focusing on clinical trial data, we sub-sampled the dataset into smaller cohorts and used the remainder as the models' evaluation dataset. Furthermore, we have changed the treatment allocation to be balanced and randomized.



\section{Metrics} \label{sect:metrics}
The evaluation of the different methods' accuracy employed two metrics, aimed at measuring different performance aspects:
\begin{itemize}
    \item \textbf{sRMSE}: scaled Root Mean Squared Error. The sRMSE is calculated as follows:   $$\mathrm{sRMSE}:=\frac{\sqrt{\sum_{i}(\hat\tau_i-\tau_i)^2 / n}} {\sqrt{\sum_{i}(\tau_i-\bar\tau)^2 / n}} = \sqrt{\frac {\sum_{i}(\hat\tau_i-\tau_i)^2} {\sum_{i}(\tau_i-\bar\tau)^2}}$$
    sRMSE measures the distance between predictions and actual CATE. Similar to RMSE, larger errors contribute substantially more to the criteria compared to smaller errors.  We scale the RMSE by the standard deviation of the CATE to obtain a scenario invariant metric that is easily interpretable and gives a sense of the relationship between the typical estimation error and the overall dynamic range of the predicted values. 
    \item \textbf{RoD}: Rate of Discordance
    $$\mathrm{RoD}:=\frac{1-\mathcal{K}(\hat\tau, \tau)}{2}$$
    Where $\mathcal{K}$ marks Kendall's rank correlation coefficient. \footnote{Kendall Tau is usually referred to as $\tau$. We denote it by $\mathcal{K}$ here to avoid confusion with the CATE.}
    We call this the rate of discordance because it is equivalent to the proportion of pairs (out of all possible pairs of observations) that are incorrectly ordered by the estimator.
\end{itemize}

In addition to calculating these metrics over individuals, we also propose an adaptation that averages errors over potential subgroups (SG). We denote a subgroup by $g_j$ which is a set of patients indices. We define $\{g_1, \ldots, 
 g_G \}$, the set of all potential subgroups of interest (for example - all half-open boxes defined by two variables and having some minimal size). For each subgroup we compute the true and estimated CATE, 
 $$\tau_{g_j}:=\frac{\sum_{i\in{g_j}} \tau_i}{|g_j|}$$
and 
\begin{equation}
\label{eq:sg_train}
\hat\tau_{g_j}:=\frac{\sum_{i\in{g_j}} y_i \cdot a_i}{\sum_{i\in{g_j}} a_i} - \frac{\sum_{i\in{g_j}} y_i \cdot (1 - a_i)}{\sum_{i\in{g_j}} (1 - a_i)}.
\end{equation} 
 
We then calculate the same accuracy metrics (sRMSE and RoD) using $\tau_{SG}$ and $\hat\tau_{SG}$ as the actual and predicted values, accordingly. E.g., 
$$\mathrm{sRMSE}_{\mathrm{SG}} = \sqrt{\frac {\sum_{j=1}^G(\hat\tau_{g_j}-\tau_{g_j})^2} {\sum_{j=1}^G(\tau_{g_j}-\bar\tau_{\mathrm{SG}})^2}}, $$ where $\bar{\tau}_{\mathrm{SG}} = \frac{1}{G} \sum_{j=1}^G \tau_{g_j}$. These metrics are more relevant for evaluating performance on a predetermined set of subgroups. CATE estimation methods which perform well on the subgroup metrics can be recommended when the aim of the CATE estimation is to serve as a basis for identifying subgroups. 

\subsection{Subgroup Definition} \label{sect:sg_def}

The set of subgroups we defined as a basis for this analysis are all 2-dimensional half-open boxes in the feature space, defined by filtering on the $\sqrt{20\%}$-th quantile of one feature to obtain a subset $S_1$, and sequentially on the $\sqrt{20\%}$-th quantile of another feature (within $S_1$). This yields (when the features are independent) SGs of 20\% of the population which are then evaluated using the different accuracy metrics.

\section{Simulation Study} \label{sect:simulation}

Our study compares the performance of various CATE estimation models in different scenarios. The treatment effect is not directly observable in real patients, since only one potential outcome can be observed per patient. Therefore, simulated data is required in order to test the performance in a valid and reliable manner.

\subsection{Methods} \label{sect:methods}

We consider a variety of Meta-Learners, including the S-Learner, X-Learner, T-Learner, and DR-Learner. The performance of Meta-Learners is strongly influenced by the base-learners used. We included a set of base-learners spanning a wide range of complexity levels:

\begin{itemize}
    \item GLMs: Regularized (Lasso and Elastic-Net) linear regression was used to estimate outcome and pseudo outcomes, and logistic regression to estimate the propensity.
    \item Accurate GLM (AGLM): This method utilizes Lasso regression to fit a piece-wise constant function, by first encoding each variable into nested bins \citep{fujita2020aglm}.
    \item Boosted Regression Trees: A sum of trees, where each tree is fitted on the residuals of the previous one \citep{chen2016xgboost}.
    \item Random Forests (RF): An average of trees constructed with stochastic sampling of features and sample to induce variability in the trees, which acts as a form of regularization \citep{breiman2001random}.
    \item Bayesian Additive Regression Trees (BART): Inspired by ensemble methods, with boosting in particular, BART also trains a sum-of-trees model, with the addition of a Bayesian regularization prior which controls the parameters of that model \citep{chipman2012bart, dorie2022stan}.
\end{itemize}
All meta-learners and ensemble methods used in the simulation use logistic regression for the estimation of the propensity score. The meta-learners and base-learners used in the simulation are: X-RF, X-BART, X-AGLM, X-Boosting, S-BART, T-Linear, and DR-RF. The mentioned base-learners were used for the estimation of the CATE and the relevant nuisance parameters. Furthermore, we compare two versions of debiased CF, with and without honesty named H-CF and CF respectively.

The ensemble methods CBA, R-Stacking, T-Stacking and Causal-Stacking utilize X-RF, X-Boosting, X-AGLM, T-Elastic-net, H-CF and CF as the base CATE estimators. 
We avoid a rigid train/validation split, and instead opted for cross-fitting the base CATE estimators, and fitting the ensemble on the entirety of the data. 
R-Stacking, T-Stacking and Causal-Stacking employ RF for estimation of the nuisance functions $m, \mu_0, \mu_1$ (with limited cross-validation for selection of the hyper-parameters). R-Stacking utilizes logistic-regression for the estimation of the propensity score.  

The estimators used for the stacked-X were Elastic-net, RF, boosting, and AGLM, which were all cross-fitted using 3-fold CV and then used as input for linear regression for the outcome on all of the training set. 
Results for additional models are available in appendix \ref{appendix:results}.

\subsection{Data Generation}

The DGPs used in the simulation study are ACIC based, the linear, slightly non-linear (NL), highly NL and the PD-L1. The DGPs are described in \cref{sect:dgp}.

All DGPs were used to generate training sets of sizes that are representative of clinical trial datasets, ranging from 100 patients to 1000 patients. The performance was then evaluated on held-out test sets of 5000 patients. The only exception to this rule are the ACIC datasets which, due to their finite sample size, were tested on the remaining part of the dataset not used for training (between 3800 and 4550 patients). Each combination of scenario and training-set size was used to generate 50 random training datasets. The results reported are based on the evaluation set. 

\subsection{Results}  \label{sect:results}

\begin{figure}[ht] \
    \centering
    \includegraphics[width=0.95\linewidth ]{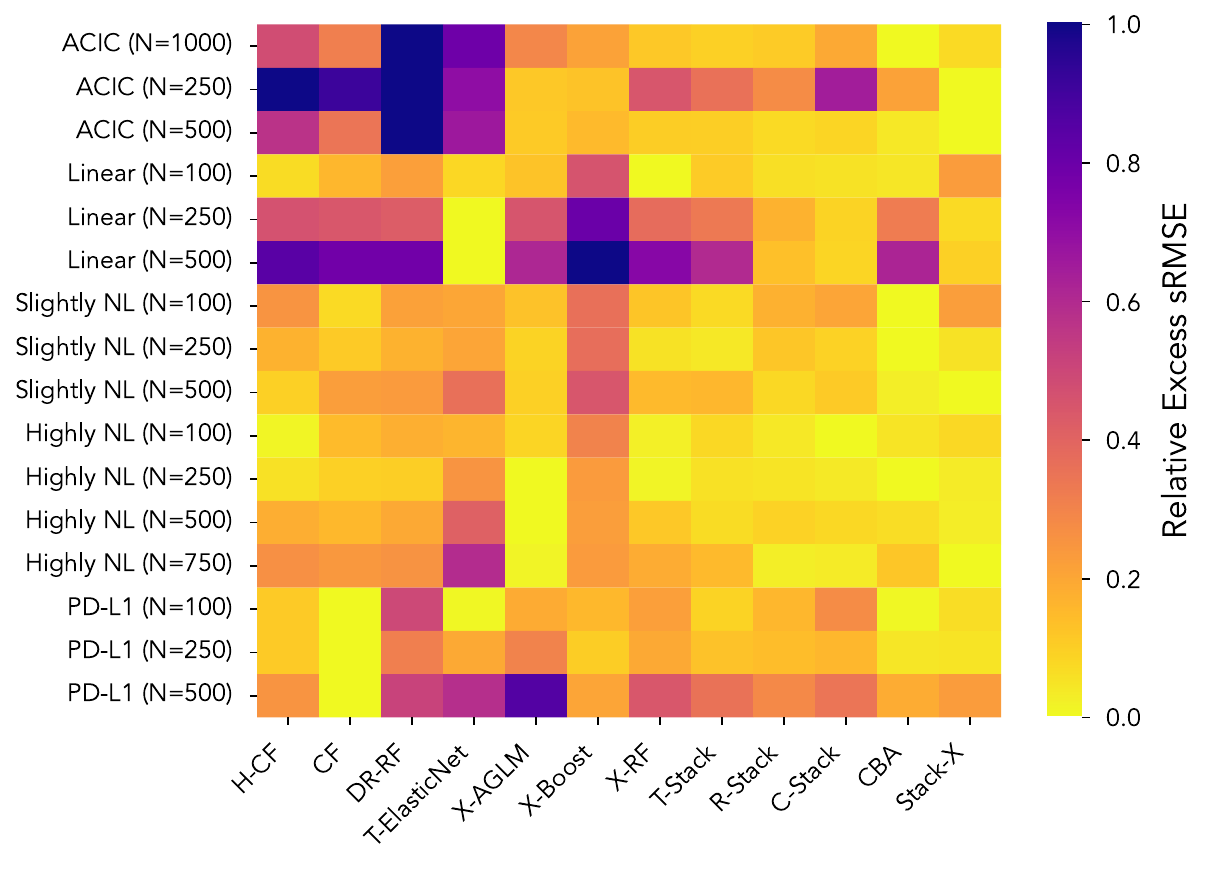}
    \captionsetup{justification=centering}
    \caption{\textbf{Relative Excess Median sRMSE}\\median sRMSE (over 50 simulation runs) was calculated for different scenarios and models. For each scenario (row) the best model was identified, and the values in the figure depict the relative excess error, compared to that of the best model.}
    \label{fig:srmse}
\end{figure}

\begin{figure}[ht]
    \centering
    \includegraphics[width=0.95\linewidth ]{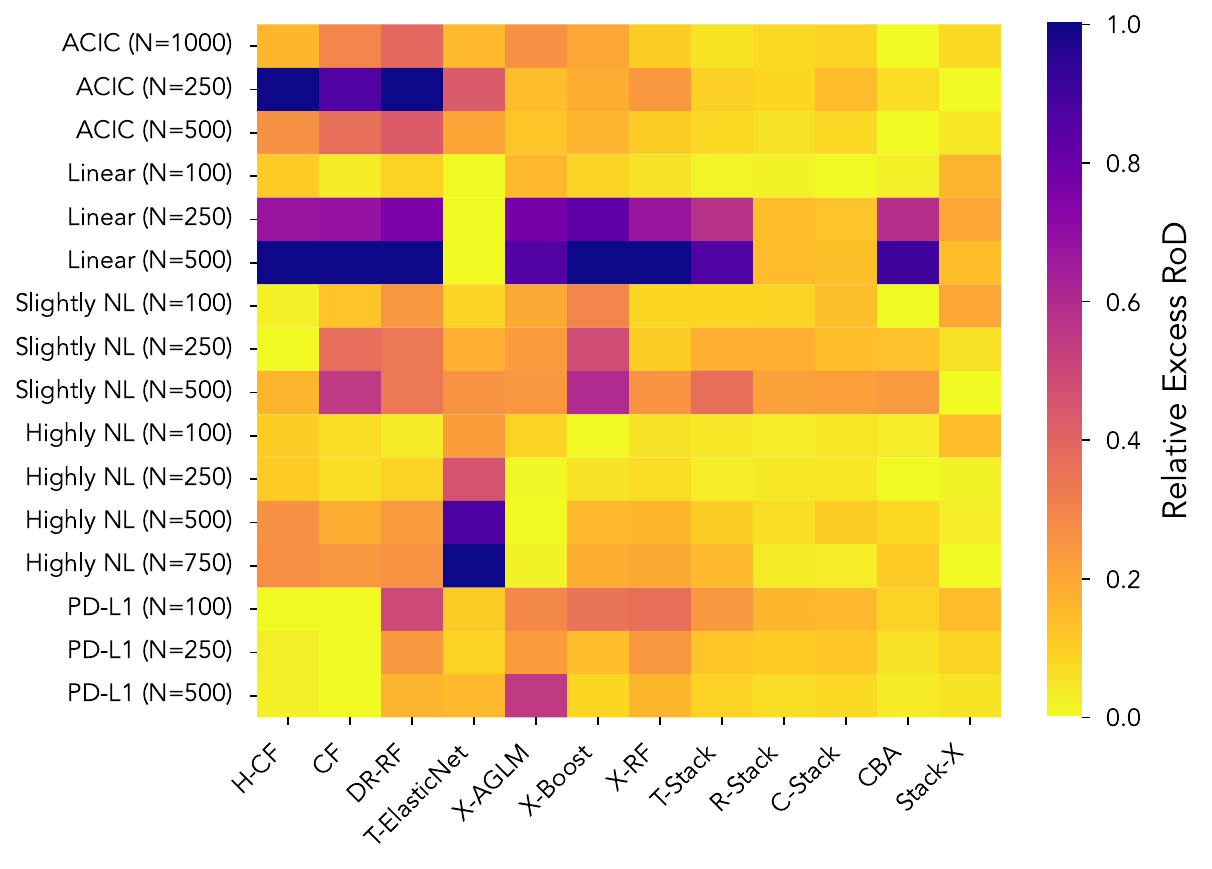}
    \captionsetup{justification=centering}
    \caption{\textbf{Relative Excess Median RoD}\\median RoD (over 50 simulation runs) was calculated for different scenarios and models. For each scenario (row) the best model was identified, and the values in the figure depict the relative excess error, compared to that of the best model.}
    \label{fig:rod}
\end{figure}

The study results demonstrate that there is no single model that is dominant across scenarios. Some models are in perfect conjunction with the DGP of some of the scenarios and therefore have an inherent advantage. For instance, the linear model-based T-Learner shows dominance in larger linear scenarios, as can be seen in Figures \ref{fig:srmse} and \ref{fig:rod}.

In general, the ensemble method appear to perform better than any single learner. Figure \ref{fig:srmse} shows the advantage of ensemble methods (CBA and Stacked-X Learner) in achieving competitive accuracy, achieving best or near-best performance in most tested scenarios. Furthermore, these ensemble methods also outperform the other ensemble approaches, R-Stacking, T-Stacking and Causal-Stacking in most scenarios. Figure \ref{fig:rod} confirms this for the RoD as well.

\FloatBarrier

\subsection{Subgroup Evaluation}
Comparing the accuracy of aggregated CATE estimation in potential subgroups supports the above findings, with the ensemble methods having good accuracy in most scenarios. Stacked-X Learner generally outperforms other ensembles.

Due to the overlap in patients between the potential subgroups as they were defined in \cref{sect:sg_def}, the estimation errors for subgroups are not independent of each other. Nevertheless, this perspective gives intuition for the expected usefulness of an estimator for subgroup identification and evaluation.

\begin{figure}[ht]
    \centering
    \includegraphics[width=0.95\linewidth ]{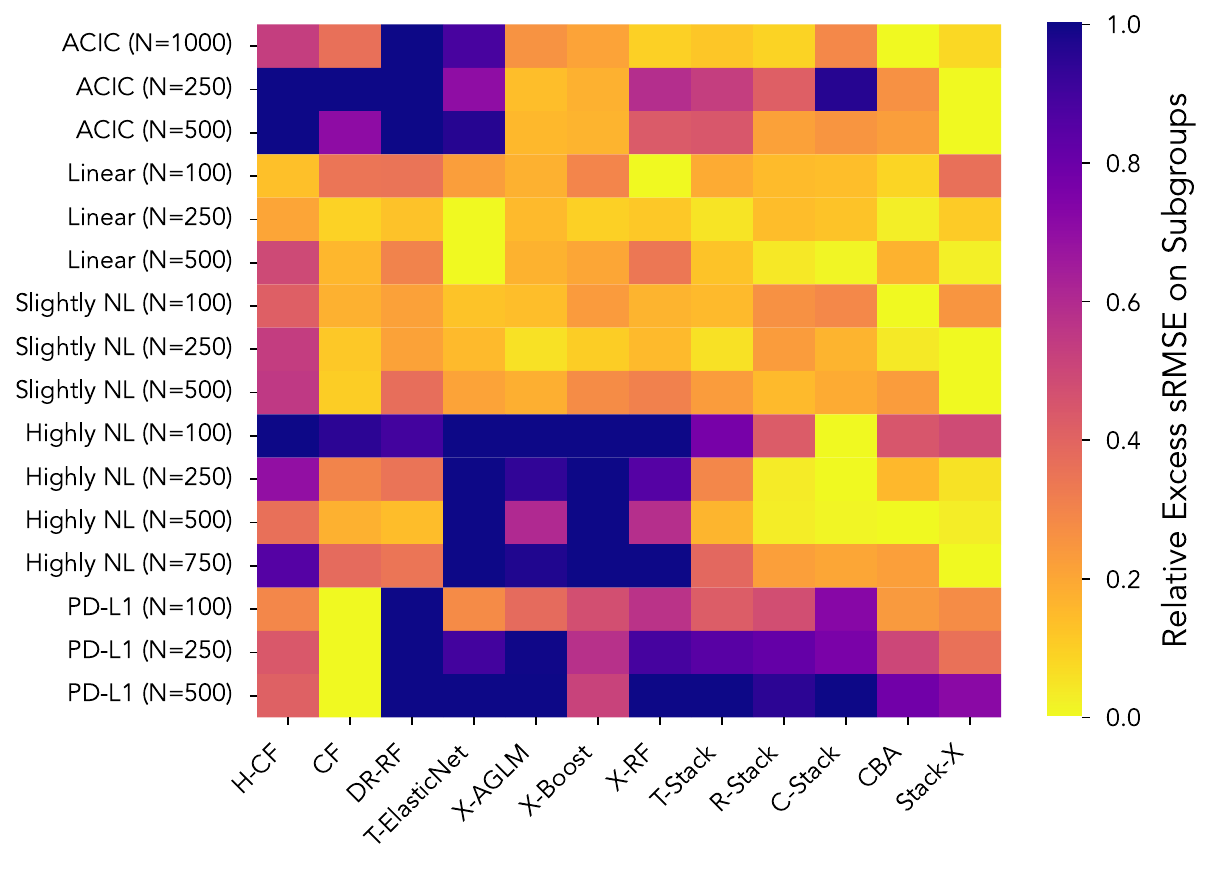}
    \captionsetup{justification=centering}
    \caption{\textbf{Relative Excess Median sRMSE on Subgroups}\\median sRMSE on subgroups (in 50 simulation runs) was calculated for different scenarios and models. For each scenario (row) the best model was identified, and the values in the figure depict the relative excess median sRMSE, compared to the best median sRMSE for that scenario.}
    \label{fig:sg_srmse}
\end{figure}

\begin{figure}[ht]
    \centering
    \includegraphics[width=0.95\linewidth ]{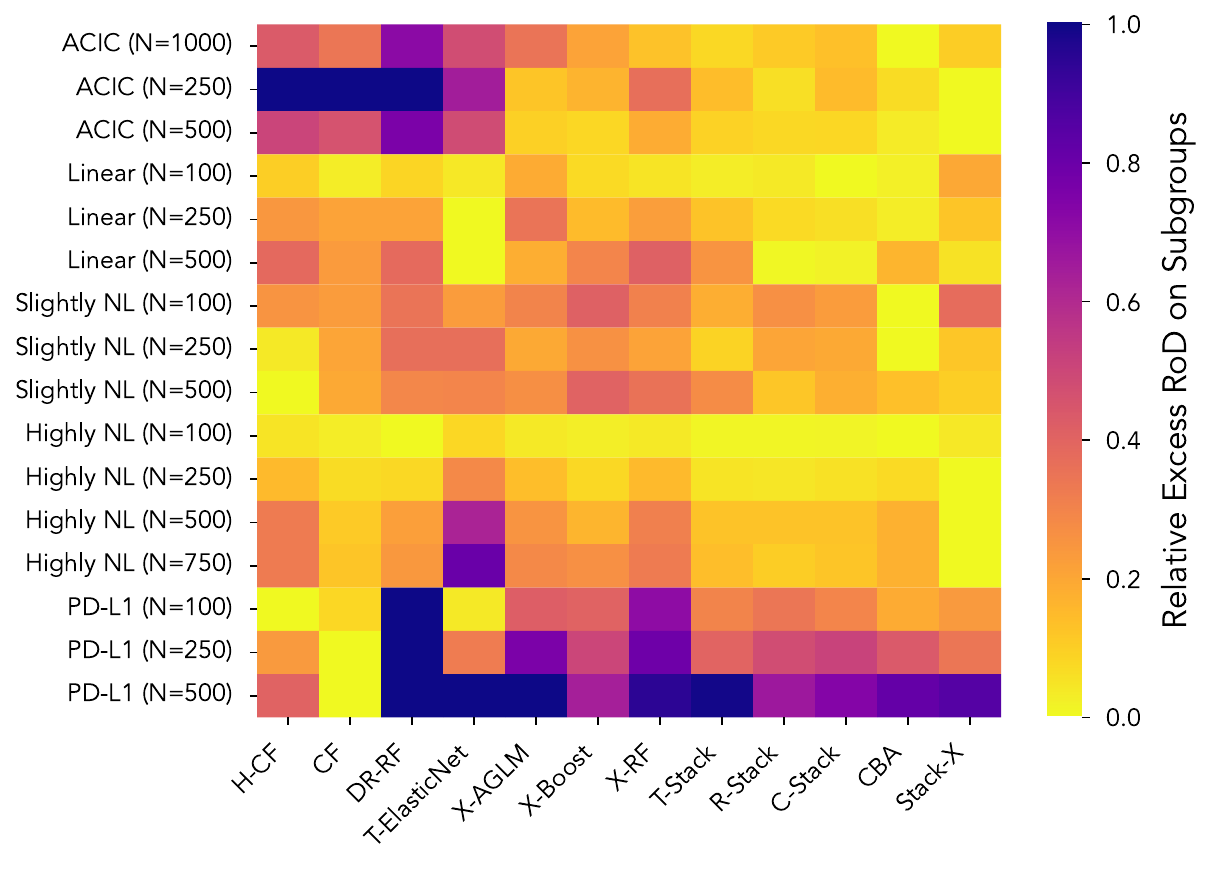}
    \captionsetup{justification=centering}
    \caption{\textbf{Relative Excess Median RoD on Subgroups}\\median RoD (in 50 simulation runs) was calculated for different scenarios and models. For each scenario (row) the best model was identified, and the values in the figure depict the relative excess median RoD, compared to the best median RoD for that scenario.}
    \label{fig:sg_rod}
\end{figure}
\FloatBarrier

\subsection{Discussion}

The results of this study highlight the robustness of ensemble methods in CATE estimation, when predicting effects at both the individual patient level and in potential subgroup evaluations. We propose ensembles of two different approaches — an X-Learner with model-stacking as a base-learner, and a Consensus-Based Average of estimators, which identifies a subset of estimators with the highest agreement.
Both ensembles approaches demonstrate superior performance over single models and ensemble models, across a wide range of scenarios. 

The stacked-X performed particularly well. We hypothesize that the performance is driven by trying to minimize losses that are observed. The rest of the CATE ensembles (T-Stacking, R-Stacking and Causal-Stacking), try to minimize losses that are dependent on some CATE estimation. Therefore, when the CATE estimation of validation set is poor, the resulting ensemble under-performs.  
Still, most ensemble methods studied consistently achieve high accuracy and exhibit robust performance across diverse scenarios compared to the single-model. 

Given the inherent variability in model performance and the unknown nature of the true DGP in practical settings, the robustness of ensemble methods is particularly valuable. This work evaluates the gain achieved by ensemble methods in setting that are relevant to clinical trial analysis - both in terms of the DGPs used and the small sample sizes used to train the models. It confirms that ensembling improves robustness also in relatively small sample sizes, despite the inherent challenge of splitting the already small dataset. This robustness is crucial for practical applications in clinical trials and personalized medicine, where accurate and stable treatment effect predictions are essential for effective decision-making.

Additional work should expand the discussion to other types of endpoints, namely binary and time-to-event endpoints, which are common in clinical research. Another question of interest is whether the advantage of Stacked X-Learner over other ensembles is preserved in other scenarios, specifically in larger data-sets.

\bibliographystyle{plainnat}
\bibliography{references}

\newpage
\begin{appendices}

\section{Consensus Based Average}
\label[appsec]{appendix:cba}
The proposed ensemble method combines predictions from multiple models by prioritizing those that exhibit higher agreement. Kendall's Tau correlation coefficients are then calculated between the predictions of each pair of models, and the mean correlation coefficient for each model is determined.

Let $\mathbf{P} = [\mathbf{p}_1, \mathbf{p}_2, \ldots, \mathbf{p}_K]$ be the matrix of predictions, where $\mathbf{p}_i$ represents the predictions from the $i$-th model, and $M$ is the number of models. Kendall's Tau correlation coefficient between models $i$ and $j$ is given by:
$$\mathcal{K}_{ij} = \mathrm{Kendall's \; Tau}(\mathbf{p}_i, \mathbf{p}_j).$$
The mean correlation coefficient for each model $i$ is:
$$\bar{\mathcal{K}}_i=\frac{1}{K-1}\sum_{j\neq i}\mathcal{K}_{ij}.$$
Next, the models are sorted in descending order based on their mean correlation values. To identify a subset of models that are in higher agreement, the method detects the "knee" in the sorted mean correlation curve, which is the point where the difference between consecutive mean correlation values is minimized. Let $\bar{\mathcal{K}}$ be the sorted mean correlation values:
$$\bar{\mathcal{K}}=[\bar{\mathcal{K}}_{(1)}, \bar{\mathcal{K}}_{(2)}, ...\bar{\mathcal{K}}_{(M)}].$$
The "knee" is detected by finding the index $m$ where the difference between consecutive mean correlation values is minimized:
$$m = \underset{x}{arg min}(\bar{\mathcal{K}}_{(i+1)} - \bar{\mathcal{K}}_{(i)}),$$
models up to this "knee" point are selected for the final ensemble. The predictions from these selected models, which have higher mean correlations and thus more agreement, are averaged to produce the final ensemble prediction. Let $\mathcal{S}$ be the set of selected models up to the knee point:
$$\mathcal{S}=\{\mathbf{p}_{(1)}, \mathbf{p}_{(2)}, \ldots,  \mathbf{p}_{(m)}\}.$$
The final ensemble prediction $\mathbf{p}_{\text{ensemble}}$ is:
$$\mathbf{p}_{\text{ensemble}} = \frac{1}{m}\sum_{\mathbf{p}_i\in S} \mathbf{p}_i.$$
By averaging the predictions from these selected models, the ensemble method leverages the consensus among models, enhancing the robustness and accuracy of the ensemble's performance.

\section{Supplementary Results}
\label[appsec]{appendix:results}

\begin{table}[!h] \label{tab:table_1_rmse}
\begin{center}
\resizebox{\textwidth}{!}{%
\begin{tabular}{ c c c | c c c c c c c c c c c }
Scenario & p & n & CF-DML & CF-DML & S-BART & T-Linear & X-AGLM & X-BART & X-Boosting & X-RF & DR-RF & Stacked-X & CBA\\\hline
\multirow{3}{*}{PD-L1} & \multirow{3}{*}{5} & 100 & 0.88 & 0.79\textbf{*} & 1.20 & 0.80 & 0.94 & 1.03 & 0.92 & 0.97 & 1.19 & 0.81 & 0.81 \\\arrayrulecolor{gray!25}\cline{3-14}\arrayrulecolor{black}
& & 250 & 0.69 & 0.62\textbf{*} & 0.98 & 0.73 & 0.80 & 0.75 & 0.68 & 0.74 & 0.81 & 0.65 & 0.66 \\\arrayrulecolor{gray!25}\cline{3-14}\arrayrulecolor{black}
& & 500 & 0.57 & 0.45\textbf{*} & 0.93 & 0.72 & 0.84 & 0.65 & 0.55 & 0.65 & 0.69 & 0.55 & 0.58 \\\arrayrulecolor{gray!25}\cline{1-14}\arrayrulecolor{black}
\multirow{4}{*}{Highly NL} & \multirow{4}{*}{10} & 100 & 1.04 & 1.18 & 1.01\textbf{*} & 1.19 & 1.11 & 1.08 & 1.34 & 1.05 & 1.21 & 1.11 & 1.05 \\\arrayrulecolor{gray!25}\cline{3-14}\arrayrulecolor{black}
& & 250 & 0.98 & 1.02 & 1.01 & 1.16 & 0.93 & 0.93\textbf{*} & 1.14 & 0.94 & 1.02 & 0.97 & 0.93 \\\arrayrulecolor{gray!25}\cline{3-14}\arrayrulecolor{black}
& & 500 & 0.90 & 0.88 & 1.01 & 1.08 & 0.77\textbf{*} & 0.82 & 0.94 & 0.85 & 0.91 & 0.78 & 0.84 \\\arrayrulecolor{gray!25}\cline{3-14}\arrayrulecolor{black}
& & 750 & 0.84 & 0.82 & 1.00 & 1.05 & 0.67 & 0.78 & 0.82 & 0.79 & 0.83 & 0.65\textbf{*} & 0.78 \\\arrayrulecolor{gray!25}\cline{1-14}\arrayrulecolor{black}
\multirow{3}{*}{Slightly NL} & \multirow{3}{*}{10} & 100 & 0.88 & 0.76 & 0.99 & 0.85 & 0.80 & 0.72 & 0.96 & 0.79 & 0.86 & 0.87 & 0.67\textbf{*} \\\arrayrulecolor{gray!25}\cline{3-14}\arrayrulecolor{black}
& & 250 & 0.61 & 0.58 & 0.95 & 0.63 & 0.57 & 0.51\textbf{*} & 0.72 & 0.55 & 0.61 & 0.54 & 0.52 \\\arrayrulecolor{gray!25}\cline{3-14}\arrayrulecolor{black}
& & 500 & 0.44 & 0.50 & 0.94 & 0.55 & 0.44 & 0.42 & 0.58 & 0.46 & 0.50 & 0.39\textbf{*} & 0.42 \\\arrayrulecolor{gray!25}\cline{1-14}\arrayrulecolor{black}
\multirow{3}{*}{Slightly NL} & \multirow{3}{*}{20} & 100 & 0.98 & 0.92 & 0.99 & 1.02 & 0.90 & 0.77\textbf{*} & 0.93 & 0.85 & 1.51 & 0.89 & 0.78 \\\arrayrulecolor{gray!25}\cline{3-14}\arrayrulecolor{black}
& & 250 & 0.76 & 0.62 & 0.98 & 0.71 & 0.61 & 0.55\textbf{*} & 0.72 & 0.72 & 0.71 & 0.58 & 0.59 \\\arrayrulecolor{gray!25}\cline{3-14}\arrayrulecolor{black}
& & 500 & 0.55 & 0.50 & 0.97 & 0.65 & 0.50 & 0.47\textbf{*} & 0.60 & 0.62 & 0.59 & 0.48 & 0.48 \\\arrayrulecolor{gray!25}\cline{1-14}\arrayrulecolor{black}
\multirow{3}{*}{Linear} & \multirow{3}{*}{10} & 100 & 0.99 & 1.07 & 1.00 & 1.00 & 1.04 & 0.94 & 1.34 & 0.92 & 1.13 & 1.09 & 0.91\textbf{*} \\\arrayrulecolor{gray!25}\cline{3-14}\arrayrulecolor{black}
& & 250 & 0.86 & 0.85 & 0.98 & 0.59\textbf{*} & 0.85 & 0.73 & 1.06 & 0.81 & 0.84 & 0.67 & 0.73 \\\arrayrulecolor{gray!25}\cline{3-14}\arrayrulecolor{black}
& & 500 & 0.76 & 0.74 & 0.97 & 0.41\textbf{*} & 0.67 & 0.59 & 0.85 & 0.71 & 0.74 & 0.46 & 0.60 \\\arrayrulecolor{gray!25}\cline{1-14}\arrayrulecolor{black}
\end{tabular}}
\caption{median sRMSE (over 50 simluated datasets)}
\end{center}
\end{table}

\begin{table}[!h] \label{tab:table_2_rod}
\begin{center}
\resizebox{\textwidth}{!}{%
\begin{tabular}{ c c c | c c c c c c c c c c c }
Scenario & p & n & CF-DML & CF-DML & S-BART & T-Linear & X-AGLM & X-BART & X-Boosting & X-RF & DR-RF & Stacked-X & CBA\\\hline
\multirow{3}{*}{PD-L1} & \multirow{3}{*}{5} & 100 & 0.25\textbf{*} & 0.25 & 0.40 & 0.28 & 0.33 & 0.32 & 0.34 & 0.35 & 0.38 & 0.29 & 0.28 \\\arrayrulecolor{gray!25}\cline{3-14}\arrayrulecolor{black}
& & 250 & 0.23 & 0.23\textbf{*} & 0.32 & 0.25 & 0.28 & 0.28 & 0.26 & 0.28 & 0.28 & 0.25 & 0.24 \\\arrayrulecolor{gray!25}\cline{3-14}\arrayrulecolor{black}
& & 500 & 0.22 & 0.21\textbf{*} & 0.30 & 0.24 & 0.33 & 0.25 & 0.23 & 0.24 & 0.25 & 0.22 & 0.22 \\\arrayrulecolor{gray!25}\cline{1-14}\arrayrulecolor{black}
\multirow{4}{*}{Highly NL} & \multirow{4}{*}{10} & 100 & 0.45 & 0.44 & 0.48 & 0.50 & 0.44 & 0.41 & 0.41\textbf{*} & 0.43 & 0.42 & 0.46 & 0.42 \\\arrayrulecolor{gray!25}\cline{3-14}\arrayrulecolor{black}
& & 250 & 0.38 & 0.36 & 0.47 & 0.50 & 0.34\textbf{*} & 0.35 & 0.36 & 0.36 & 0.37 & 0.36 & 0.35 \\\arrayrulecolor{gray!25}\cline{3-14}\arrayrulecolor{black}
& & 500 & 0.34 & 0.32 & 0.47 & 0.50 & 0.27\textbf{*} & 0.29 & 0.31 & 0.31 & 0.33 & 0.27 & 0.30 \\\arrayrulecolor{gray!25}\cline{3-14}\arrayrulecolor{black}
& & 750 & 0.30 & 0.29 & 0.47 & 0.50 & 0.24 & 0.27 & 0.28 & 0.28 & 0.29 & 0.23\textbf{*} & 0.27 \\\arrayrulecolor{gray!25}\cline{1-14}\arrayrulecolor{black}
\multirow{3}{*}{Slightly NL} & \multirow{3}{*}{10} & 100 & 0.21 & 0.24 & 0.38 & 0.23 & 0.25 & 0.22 & 0.27 & 0.23 & 0.26 & 0.25 & 0.20\textbf{*} \\\arrayrulecolor{gray!25}\cline{3-14}\arrayrulecolor{black}
& & 250 & 0.15 & 0.21 & 0.32 & 0.18 & 0.19 & 0.16 & 0.23 & 0.17 & 0.20 & 0.15 & 0.15\textbf{*} \\\arrayrulecolor{gray!25}\cline{3-14}\arrayrulecolor{black}
& & 500 & 0.14 & 0.19 & 0.32 & 0.16 & 0.15 & 0.14 & 0.20 & 0.16 & 0.17 & 0.13\textbf{*} & 0.14 \\\arrayrulecolor{gray!25}\cline{1-14}\arrayrulecolor{black}
\multirow{3}{*}{Slightly NL} & \multirow{3}{*}{20} & 100 & 0.33 & 0.31 & 0.41 & 0.29 & 0.28 & 0.25 & 0.28 & 0.27 & 0.34 & 0.26 & 0.25\textbf{*} \\\arrayrulecolor{gray!25}\cline{3-14}\arrayrulecolor{black}
& & 250 & 0.19 & 0.22 & 0.39 & 0.22 & 0.21 & 0.18 & 0.24 & 0.22 & 0.25 & 0.18 & 0.18\textbf{*} \\\arrayrulecolor{gray!25}\cline{3-14}\arrayrulecolor{black}
& & 500 & 0.16 & 0.19 & 0.39 & 0.20 & 0.17 & 0.15 & 0.20 & 0.18 & 0.19 & 0.15 & 0.15\textbf{*} \\\arrayrulecolor{gray!25}\cline{1-14}\arrayrulecolor{black}
\multirow{3}{*}{Linear} & \multirow{3}{*}{10} & 100 & 0.35 & 0.33 & 0.42 & 0.32 & 0.37 & 0.30\textbf{*} & 0.34 & 0.33 & 0.35 & 0.36 & 0.32 \\\arrayrulecolor{gray!25}\cline{3-14}\arrayrulecolor{black}
& & 250 & 0.26 & 0.27 & 0.39 & 0.16\textbf{*} & 0.28 & 0.23 & 0.29 & 0.26 & 0.28 & 0.20 & 0.22 \\\arrayrulecolor{gray!25}\cline{3-14}\arrayrulecolor{black}
& & 500 & 0.23 & 0.23 & 0.39 & 0.11\textbf{*} & 0.21 & 0.18 & 0.25 & 0.23 & 0.24 & 0.13 & 0.17 \\\arrayrulecolor{gray!25}\cline{1-14}\arrayrulecolor{black}
\end{tabular}}
\caption{median RoD (over 50 simulated datasets)}
\end{center}
\end{table}

\end{appendices}

\end{document}